\documentclass[letter,11pt]{article}
\usepackage[margin=2cm]{geometry}
\usepackage[sumlimits,intlimits]{amsmath}
\usepackage{amsfonts,amssymb}
\usepackage{mathrsfs}
\usepackage{amsmath}
\usepackage[T1]{fontenc}
\usepackage{capt-of}
\usepackage{ucs}
\usepackage{graphics}
\usepackage[dvips]{graphicx}
\usepackage{textcomp}
\usepackage{caption}

\DeclareGraphicsExtensions{.pdf,.png,.eps}
\usepackage{lineno}
\usepackage[square,sort&compress,comma]{natbib}
\setcitestyle{numbers} 
\newenvironment{mylisting}

\begin{document}
\makeatletter
\@addtoreset{equation}{section}
\makeatother
\renewcommand{\theequation}{\thesection.\arabic{equation}}


\vspace{.5cm}
\begin{center}
\Large{\bf  More stable dS vacua from S-dual non-geometric fluxes}\\

\vspace{1cm}

\large
Cesar Damian\footnote{cesaredas@fisica.ugto.mx} and  Oscar Loaiza-Brito\footnote{oloaiza@fisica.ugto.mx} \\[4mm] 
{\small\em Departamento de F\'isica, DCI,  Campus Le\'on,}\\
{\small\em Universidad de Guanajuato, C.P. 37150, Leon, Guanajuato, Mexico.}\\[4mm]
\vspace*{2cm}
\small{\bf Abstract} \\
\end{center}

\begin{center} 
\begin{minipage}[h]{14.0cm} { Stable vacua obtained from isotropic tori compactification might not be fully stable  provided the existence of runaway directions in the K\"ahler directions of anisotropy. By implementing a genetic algorithm we report  the existence of explicit flux configurations leading to stable de Sitter and Anti- de Sitter vacua, consisting on Type IIB compactifications on a 6-dimensional anisotropic torus threaded with standard and S-dual invariant non-geometric fluxes in the presence of orientifold 3-planes. In all dS vacua  the masses of the complex structure moduli are heavier than the Hubble scale suggesting that the axio-dilaton and K\"ahler moduli are natural candidates for small-field inflation.  In the way, we also report new solutions on isotropic and semi-isotropic tori compactifications. Finally, we observe that, since all our solutions are obtained in the absence of solitonic objects, they are good candidates to be lifted to stable solutions  in extended supersymmetric theories.}

\end{minipage} 
\end{center}
 
%


\newpage

\section{Introduction}
Over the last few years, the search for (meta)stable vacua has been the center of great interest by the string cosmology community \cite{Kachru:2003aw, Kallosh:2007ig} (for a recent review see \cite{Andriot:2013txa} and references therein). In particular, there is a  class of models which has been studied in detailed concerning a Type II compactification on an (an)isotropic  torus in the presence of a $\mathbf{Z}_2\times\mathbf{Z}_2$ orientifold. 
Within these models,  the existence of dS vacua has been analyzed by the use of different methods as algebraic geometry, direct numerical calculations and statistical analisys \cite{Danielsson:2009ff, Danielsson:2010bc, Chen:2011ac,Shiu:2011zt,Marsh:2011aa,Danielsson:2011au,Caviezel:2009tu}. In the context of Type IIA with pure geometrical fluxes, dS vacua turn out to be unstable since the presence of tachyonic directions seems to be a generic feature in these models\cite{ Dibitetto:2010rg, Borghese:2011en,Caviezel:2008tf,Flauger:2008ad}.  On the other hand, stable dS solutions in type II theories in the presence of non-geometric fluxes and orbifold singularities have been reported recently \cite{Dibitetto:2012ia, Danielsson:2012by, Danielsson:2012et, Blaaback:2013ht}. In particular,  the authors in \cite{Dibitetto:2011gm} show some solutions which although stable in the context of ${\cal N}=1$ supergravity, become unstable by considering the full set of scalar fields in  ${\cal N}=4$ supergravity.\\

In this work we continue our search,  initiated in \cite{Damian:2013dq},  for stable de Sitter (dS) and anti-de Sitter (AdS) vacua by implementing a genetic algorithm in a type IIB compactification on a 6-dimensional factorizable torus threaded with  Ramond-Ramond (RR), Neveu-Schawrz-Neveu-Schwarz (NS-NS) and S-dual non-geometric fluxes denoted $(Q,P)$ \cite{Wecht:2007wu, Aldazabal:2006up} in the presence of orientifold 3-planes.  Hence, the four-dimensional effective theory under consideration is the ${\cal N}=4$ Type IIB $T^6/\mathbf{Z}_2$ orientifold (originally analyzed in \cite{Frey:2002hf, Kachru:2002he} and generalized  to non-geometric fluxes in \cite{Flournoy:2004vn, Shelton:2005cf, Shelton:2006fd, Wecht:2007wu, Aldazabal:2006up}). In this scenario, the orientifold acts through the involution $\Omega\cdot {\cal I}_6\cdot (-1)^{F_L}$, where $\Omega$ inverts the string world-sheet orientation and ${\cal I}_6$ inverts the entire $T^6$ implying the presence of 38 real moduli out of which, 21 of them correspond to the metric moduli (complex structure), 15 are K\"ahler moduli and 2 real moduli related to the complex axio-dilaton.  The present work focus on a subset of this  moduli space by considering a factorizable torus $T^6$, which correspoinds to the same subset as that of the untwisted sector of the ${\cal N}=1$ Type IIB orientifold of the Calabi-Yau orbifold $\mathbf{Z}_2\times\mathbf{Z}_2$.  Actually we are focusing in the most general case (an anisotropic factorizable torus) on a subset consisting on 6 metric moduli, 6 K\"ahler moduli and 2 real moduli for the axio-dilaton. Therefore, there could be runaway directions represented by moduli in the K\"ahler potential which do not appear in the effective superpotential.  Considering the whole moduli space is undoubtedly the next step to implement towards the construction of fully stable vacua. We leave this important case for future work.\\ 

In this work we concentrate our analysis on the factorizable torus,  having in mind that our goal is twofold: 1) Stable vacua constructed by compactification on isotropic torus could be unstable provided the existence of runaway directions in the K\"ahler directions of anisotropy\footnote{We thank R. Blumenhagen for comments regarding this point.}. Therefore, we want to find explicit examples of stable vacua obtained by compactifications on anisotropic tori, and 2) we are interested in studying how the presence of extra fluxes,  as the $P$-fluxes, in isotropic, semi-isotropic or anisotropic torus compactifications, alters (or not) some generic cosmological features of the model, as the presence of sub-Hubble massive moduli.\\

With the purpose of simplifying our system we also assume the following:
1)There is not a priori selection of which  moduli break SUSY, 2) there are not exotic orientifolds (all fluxes are taken even) and  3) there are not sources for the considered fluxes.  It is possible that our solutions could be lifted to stable solutions in a ${\cal N}=4$ supergravity, although deeper studies are required for a decisive conclusion\footnote{We thank A. Guarino for explaining us this.}.\\
%

Henceforth, with all these considerations in mind, we have applied our genetic algorithm for searching for stable vacua. In consequence, we report  11 AdS and  7 dS stable vacua. Although we are still far to present a full analysis, we find some generic features within these explicit models: 
\begin{enumerate}
\item
There is always a mass hierarchy on the moduli mass. It seems that (some of the) complex structure moduli are heavier than the rest. 
\item
Provided there are sub-Hubble massive moduli,  it seems that  (most of them) correspond to the K\"ahler moduli,  favoring a scenario in which K\"ahler moduli might drive multi-field inflation in the small field regime.
\item
For the reported solutions, SUSY breaking scales, masses of gravitinos and vacuum energy values (vev's) are still to high for phenomenological purposes and there is not a clear correlation among those values and the presence of $Q$ or $P$ fluxes, or if the vacua is constructed by compactification on an isotropic, semi-isotropic or anisotropic torus.
\end{enumerate}

We have also find some interesting features on some solutions as:
\begin{enumerate}
\item
We report a supersymmetric stable AdS solution without imposing supersymmetry.
\item
We report the existence of two stable dS solutions in an anisotropic torus although within the BF limit (see below).
\end{enumerate}

It is important to remark that our search is limited to some ad hoc selection of families of solutions in the flux configuration space which does not allow us to search the full space of flux configurations. Therefore, although the above features seems to be generic, they are the result of some concrete solutions we have found. A more complete analysis is still under work.\\

Equally important is to notice that all these 4-dimensional models constructed from the inclusion of S-dual non-geometric fluxes are still not derived from a ten-dimensional string theory. Even more, it is still unclear how to lift generic non-geometric flux configurations to (some recent attempts of) a 10-dimensional bi-invariant theory,  although huge efforts  have been made over the last few years. Henceforth, it would be interesting to see whether our specific models could be obtained from a compactification of a ten-dimensional theory \cite{Andriot:2011uh, Andriot:2010ju} or within the context of double field theory (for a recent work see \cite{Geissbuhler:2013uka}).

\section{The effective scalar potential}
Following the analysis performed in \cite{Damian:2013dq} we look for stable de Sitter (dS) and Anti de
Sitter (AdS) vacua in the context of Type IIB superstring compactification on an anisotropic torus threaded with non-geometric fluxes  $Q$ {\it and their S-dual counterparts} referred as $P$-fluxes. Standard orientifold three-planes act by a $\mathbf{Z}_2$ symmetry on the internal space, i.e., we do not consider the presence of $O7$-planes, $D$-branes or induced $I7$-branes. The effective supersymmetric theory contains seven complex moduli fields, denoted by $\Phi_i=\phi_i+i\psi=\{\tau_{1,2,3}, S, U_{1,2,3}\}$, where $\tau_i$ is the complex structure, $S$ the axio-dilaton and $U_i$ the K\"ahler moduli. The corresponding superpotential is given by \cite{Aldazabal:2006up}
\begin{equation}
W=P_0-\Phi_4 P_1 +\sum_{m=5}^{7}\left(P_2^m-\Phi_4  P_3^m\right)\Phi_m,
\end{equation}
where $P_i =P_i(\Phi_1, \Phi_2, \Phi_3)$ are polynomial of cubic order on the complex structure with real coefficients as follows (where the convention of summing on contracted indices is taken):
\begin{eqnarray}
P_0(\Phi_{1,2,3})&=&a_{00}^0-a_{01}^i\Phi_i+ a_{02}^{ij}\Phi_i\Phi_j-a^{ijk}_{03}\Phi_i\Phi_j\Phi_k,\nonumber\\
P_1(\Phi_{1,2,3})&=&a_{10}^0-a_{11}^i\Phi_i+ a_{12}^{ij}\Phi_i\Phi_j-a^{ijk}_{13}\Phi_i\Phi_j\Phi_k,\nonumber\\
P_2^m(\Phi_{1,2,3})&=&a_{20}^{m0}+a_{21}^{mi}\Phi_i- a_{22}^{mij}\Phi_i\Phi_j-a^{mijk}_{23}\Phi_i\Phi_j\Phi_k,\nonumber\\
P_3^m(\Phi_{1,2,3})&=&a_{30}^{m0}+a_{31}^{mi}\Phi_i+ a_{32}^{mij}\Phi_i\Phi_j+a^{mijk}_{33}\Phi_i\Phi_j\Phi_k,
\end{eqnarray}
where $a^{mijk}_{rs}$ correspond to RR  ($r=0$), NS-NS ($r=1$), Q (r=2) and P-fluxes ($r=3$) integrated over 3-cycles and with one leg over each torus, and the index $m$ runs from 1 to 3. Notice that for the isotropic case the upper indices $i,j,k$ are all equal since $\Phi_1=\Phi_2=\Phi_3$, while in the anisotropic torus, all of them are different. See Table \ref{coeff} for our notation.\footnote{Only invariant fluxes to the orientifold projection are shown. All terms related to $H_3$, $Q$ and $P$-fluxes are also multiplied by the axiodilaton $\Phi_4$.}\\

\begin{center}
\begin{table}[h]
\begin{minipage}[t]{3.90in}
\caption{\label{coeff} Integrated fluxes. Latin indices correspond to  horizontal coordinates on the torus, while greek indices are related to vertical coordinates.}
\end{minipage}
\centering
\begin{tabular}{@{}*{7}{l}}
\hline
Term in $W$ &IIB Flux & $a^{mijk}_{rs}$\\
\hline
$1$					&$\bar{F}_{ijk}$														&$a_{00}^{0}$\\
$\Phi_i$				&$\bar{F}_{ij\gamma}$ 												& $a_{01}^{i}$\\
$\Phi_i \Phi_j$			&$\bar{F}_{i{\beta}\gamma}$ 											& $a_{02}^{ij}$\\
$\Phi_i \Phi_j \Phi_k$	&$\bar{F}_{\alpha{\beta}\gamma}$ 										& $a_{03}^{ijk}$\\
$1$					&$\bar{H}_{ijk}$ 													& $a_{10}^{0}$\\
$\Phi_i$				&$\bar{H}_{\alpha{jk}}$ 												& $a_{11}^{i}$\\
$\Phi_i \Phi_j$			&$\bar{H}_{i\beta\gamma}$ 											& $a_{12}^{ij}$\\
$\Phi_i \Phi_j \Phi_k$	&$\bar{H}_{\alpha\beta\gamma}$ 										& $a_{13}^{ijk}$\\
$1$					&$\bar{Q}_{k}^{\alpha\beta}$ 											& $a_{20}^{0}$\\
$\Phi_i$				&$\bar{Q}_{k}^{\alpha{j}}$,$\bar{Q}_{k}^{i\beta}$, $\bar{Q}_{\alpha}^{\beta\gamma}$ 	& $a_{21}^{m1}$,$a_{21}^{m2}$,$a_{21}^{m3}$\\
$\Phi_i \Phi_j$			&$\bar{Q}_{\gamma}^{i\beta}$,$\bar{Q}_{\beta}^{\gamma{i}}$,$\bar{Q}_{k}^{ij}$ 		& $a_{22}^{mi1}$,$a_{22}^{mi2}$,$a_{22}^{mi3}$\\
$\Phi_i \Phi_j \Phi_k$	&$\bar{Q}_{\gamma}^{ij}$ 												& $a_{23}^{m123}$\\
$1$					&$\bar{P}_{k}^{\alpha\beta}$ 											& $a_{30}^{0}$\\
$\Phi_i$				&$\bar{P}_{k}^{\alpha{j}}$,$\bar{P}_{k}^{i\beta}$, $\bar{P}_{\alpha}^{\beta\gamma}$ 	& $a_{31}^{m1}$,$a_{31}^{m2}$,$a_{31}^{m3}$\\
$\Phi_i \Phi_j$			&$\bar{P}_{\gamma}^{i\beta}$,$\bar{P}_{\beta}^{\gamma{i}}$,$\bar{P}_{k}^{ij}$ 		& $a_{32}^{mi1}$,$a_{32}^{mi2}$,$a_{32}^{mi3}$\\
$\Phi_i \Phi_j \Phi_k$	&$\bar{P}_{\gamma}^{ij}$ 												& $a_{33}^{m123}$\\
\hline
\end{tabular}
\end{table}
\end{center}

Substituting the above polynomials in the superpotential $W$, we are able to compute the corresponding scalar potential  given by (in Planck units)
\begin{equation}
V=e^{K} \left({\mathcal{G}^{i\bar{j}}\mathcal{G}_{i}\mathcal{G}_{\bar{j}}-3}\right)
\end{equation}
where as usual $\mathcal{G}^{i\bar{j}}$ is the inverse of  the K\"ahler metric, $\mathcal{G}{\equiv}=K-\ln (|W|^2)$  and the K\"ahler potential $K$ is
\begin{equation}
K=-\sum_{i=1}^{7} \ln (2\psi_i).
\end{equation}
The result is a scalar potential which is written as a function of  the fluxes $a^{mikj}_{rs}$ and 7 complex fields.
In principle there is a huge number of flux configurations for which the scalar potential could have a stable minimum. However, the set of fluxes are constrained by tadpole conditions and Bianchi identities. Therefore, it is necessary  to find a set of fluxes which fulfill these constraints before looking for extrema of the scalar potential.

\subsection{Constraints}
As it is known, there are basically two types of constraints a set of fluxes must fulfill. The first concerns the Tadpole conditions or roughly speaking the cancellation of D3-brane charges on $\widetilde{T}^6$:
\begin{equation}
\int_{\mathbf{\widetilde{T}}^6}H_3 \wedge F_3=16,
\end{equation}
and the corresponding constraint on the $Q$ and $P$ -fluxes by imposing $SL(2,\mathbb{Z})$ invariance and T-duality:
\begin{equation}
\int Q{\cdot}F_{3}=\int P{\cdot}H_{3}=\int Q{\cdot}H_3+P{\cdot}F_3=0,
\end{equation}
where as usual the $Q$  and $P$ fluxes contracts with a $p$-form to give a $p-1$ form. Those constraints implies the absence of sources as $D7$-branes, $NS7$-branes and the induced $I7$-branes respectively in accordance with our initial setup\footnote{We thank A. Guarino for a useful explanation about the importance of fulfilling tadpole and Bianchi constraints in the absence of sources and the lifting of those solutions to ${\cal N}=4$ supergravity.}.\\

Similarly, the dual counterpart of Bianchi identity on $H_3$ given by $dH_3=0$ is obtained by extending the algebra of isometry generators of the twisted background,  where one can observe that the  structure constants of the Lie algebra are in fact the non-geometric fluxes $Q$ and $P$.  The generalized identities are then given by (see \cite{Wecht:2007wu} for a formal derivation):
\begin{equation}
Q_{[L}^{RP}H_{MN]P}-P_{[L}^{RP}F_{MN]P}=0,{\quad}Q^{[MN}_{P}Q^{L]P}_{R}=0
\end{equation}
and
\begin{equation}
Q_{P}^{[MN}P_{R}^{L]P}+P_{P}^{[MN}Q_{R}^{L]P}=0,{\quad}P^{[MN}_{P}P^{L]P}_{R}=0.\\
\end{equation}
Notice that besides the assumption of the absence of sources, cancelation of tadpoles and Bianchi identities, $Q\cdot H=P\cdot F=0$ are fulfilled in a stepwise manner .  Regarding this, it has been observed \cite{deCarlos:2009fq, deCarlos:2009qm} that 
one way to construct stable solutions in ${\cal N}=4$ supergravity consists on finding stable solutions in ${\cal N}=1$ and uplifting. It seems that the absence of sources could increase the chances to obtain stable solutions.  As commented in \cite{Dibitetto:2011gm} there are no known stable dS solutions in ${\cal N}=4$ supergravity.  Notice however, that stable dS solutions with the above stepwise manner to cancel tadpoles and satisfy Bianchi identities is not necessarily a stable solution in ${\cal N}=4$ since stability against the rest of the scalars must still be studied. \\

\section{Stable dS and AdS}
By implementing a genetic algorithm we look for stable minima for the  scalar potential generated by a set of fluxes satisfying the above constraints.  However, we do not impose the extra constraint concerning preservation of supersymmetry by one or more moduli, which leads to the possibility to find both, supersymmetric and non-supersymmetric stable vacua.  Although the later is clearly more probable we report a flux configuration upon which a stable supersymmetric  AdS vacuum is constructed. \\

We focus our study in compactifications on isotropic, semi-isotropic (i.e., where only two torus are identical) and  anisotropic tori threaded with $Q$ and/or $P$-fluxes\footnote{Isotropic compactifications with $Q$-fluxes in the absence of $P$-fluxes were already studied in \cite{Damian:2013dq}.} and such selection is put by hand  in our algorithm at the beginning of our calculations. See Appendix B for the  algorithm  code. In all cases, we take into account NS-NS and R-R 3-form fluxes since their presence is fundamental for finding stable vacua.\\

Before we present our results, let us fix our notation concerning the label we assign to each of the cases we report. We have use a nomenclature consisting on 3 letters and a number (if the case), as follows:
\begin{itemize}
\item
For an AdS vacuum we denote the solution by an {\bf A}, while for a dS solution we use {\bf dS}.
\item
For a compactification on an isotropic, semi-isotropic or anisotropic torus, we denote the solution by {\bf I}, {\bf S} and {\bf A}, respectively.
\item
According to the non-geometric fluxes we have turned on, we denote the solution by ${\bf Q}$, ${\bf P}$ or ${\bf QP}$ where the last denotes that both fluxes are present.
\item
Finally, we use numbers to distinguish different cases which carry the same previous labels.
\end{itemize}
Therefore, if for instance, we have a dS solution by a compactification on an anisotropic torus in the presence of $Q$ and $P$-fluxes, we refer to it as the {\bf (dS-A-QP)} case.

\subsection{AdS vacua}
We report  the existence of at least 11 AdS stable vacua constructed from different flux configurations by compactifying on isotropic, semi-isotropic and anisotropic tori. Specifically: 
\begin{enumerate}
\item
For the {\it isotropic} case, we report 6 different stable vacua. In the first three cases there are not $Q$-fluxes, i.e., we are considering a compactification with only $P$-fluxes while the last three cases consider the presence of  both non-geometric fluxes $P$ and $Q$.
\item
For the {\it semi-isotropic} cases, we find 3 different configurations leading to stable vacua. We report three different scenarios where, besides the standard NS-NS and R-R fluxes, we have only $Q$, or $P$ or both non-geometric fluxes.  
\item
For the {\it anisotropic} case, we report 2 different cases concerning the presence of $Q$-fluxes and both $Q$ and $P$, but we were not able to find a flux configuration in which we have only $P$-fluxes. Notice that strictly speaking, these cases are fully stable since there are not possible run aways directions by perturbing the torus isotropy\footnote{Although we are still considering a factorizable torus. In such case, perturbations on K\"ahler and complex structure moduli away from the factorizable case must be studied as well. Compactifications on non-factorizable torus are left for future work.}
\end{enumerate}

The flux configuration leading to stable vacua for the above 3 cases are shown in Tables \ref{tab:fluxes}, \ref{tab:sifluxes} and \ref{tab:afluxes}.  It is worth to compare the vev of these solutions with the corresponding values in the  A-I-Q cases\footnote{Explicit representative solutions for AIQ cases are shown in Appendix B for completeness.} found in \cite{Damian:2013dq} which are of the order -$10^{-2}$ and $-10^{-7}$ in Planck units.  So far, among the stable cases  here reported, those obtained by compactification on isotropic torus with only $Q$ fluxes,  contain the smallest vevs.\\

\begin{table}[htdp]
\caption{\label{tab:fluxes} Flux configuration for AdS vacua. All cases correspond to compactification on an isotropic torus. The value for the scalar potential at its minimum is given in Planck units.}
\begin{center}
\begin{tabular}{c|c|cccccc}
\hline
&Flux 			& A-I-P1		  & A-I-P2 		& A-I-P3		& A-I-QP1		& A-I-QP2 		& A-I-QP3		\\
\hline
RR&$a_{00}^0$	& $4$		& $8$		& $10$		& $0$		& $0$		& $4$		\\
&$a_{01}$		& $10$ 		& $10$		& $12$ 		& $0$		& $0$		& $(12,12,12)$	\\
&$a_{02}$		& $22$ 		& $14$		& $16$		& $16$		& $8$		& $24$		\\	
&$a_{03}$		& $46$		& $22$		& $24$		& $16$		& $8$		& $64$		\\

\hline
NS-NS&$a_{10}^0$	& $10$		& $52$		& $48$		& $-2$		& $-4$		& $-2$		\\		
&$a_{11}$			& $8$		& $4$		& $2$		& $0$		& $0$		& $0$		\\
&$a_{12}$		& $-6$		& $-40$		& $-48$		& $42$		& $42$		& $0$		\\
&$a_{13}$		& $-54$		& $-24$		& $-64$		& $64$		& $46$		& $-24$		\\
\hline
$Q$&$a_{20}$		&$0$			&$0$			&$0$			& $0$		& $0$		& $0$		\\	
&$a_{21}^j$		&$(0,0,0)$		&$(0,0,0)$		&$(0,0,0)$		& $(0,0,0)$	& $(0,0,0)$	& $(-4,-4,-4)$	\\
&$a_{22}^j$		&$(0,0,0)$		&$(0,0,0)$		&$(0,0,0)$		& $(32,32,32)$	& $(8,8,8)$	& $(0,0,0)$	\\
&$a_{23}$		&$0$			&$0$			&$0$			& $32$		& $8$		& $24$		\\
\hline

$P$&$a_{30}$		& $-12$		& $-16$		& $-16$		& $-4$		& $-4$		& $0$		\\	
&$a_{31}^j$		& $(24,24,-12)$& $(32,32,-16)$& $(32,32,-16)$	& $(4,4,0)$	& $(4,4,0)$	& $(0,0,0)$	\\
&$a_{32}^j$		& $(24,24,-48)$	& $(32,32,-64)$& $(32,32,-64)$	& $(40,40,36)$	& $(38,38,34)$	& $(0,0,24)$	\\
&$a_{33}$		& $48$		& $64$		& $64$		& $40$		& $38$		& $0$		\\
\hline\hline
$V_{min}$ &		& $-548.06$	& $-24.371$	& $-13.212$	& $-19.123$	& $-0.83514$	& $-59.0505$		\\
\hline
\end{tabular}
\end{center}
\end{table}%
\begin{table}[htdp]
\caption{\label{tab:sifluxes} Flux configuration for AdS vacua from a semi-isotropic torus compactification.}
\begin{center}
\begin{tabular}{c|c|ccc}
\hline
Sector &Flux 		& A-S-Q		& A-S-P		& A-S-QP			\\
\hline
RR & $a_{000}$	& $-46$		& $0$		& $0$	\\
& $a_{01i}$		& $(14,14,14)$	& $(0,0,0)$	& $(0,0,2)$	\\
& $a_{02i}$		& $(32,32,32)$	& $(0,0,-2)$	& $(0,0,0)$\\
& $a_{033}$		& $-46$		& $-2$		& $0$				\\
\hline
NSNS & $a_{100}$	& $0$		& $32$		& $34$				\\	
& $a_{11i}$		& $(0,0,40)$	& $(2,2,16)$	& $(16,16,14)$		\\
& $a_{12i}$		& $(-32,-32,0)$	& $(-50,-50,50)$& $(-16,-16,16)$	\\
& $a_{133}$		& $48$		& $26$		& $16$		\\
\hline
Q & $a_{2i0}$		& $(0,0,24)$	& $(0,0,0)$	& $(0,0,16)$	\\	
& $a_{21i}^j$		& $\left(\begin{array}{ccc}  8 & 8  & 8  \\ 8  & 8  & 8  \\ 8  & 8  & -16  \end{array}\right)$	& 
				$\left(\begin{array}{ccc}  0 & 0  & 0  \\ 0  & 0  & 0  \\ 0  & 0  & 0  \end{array}\right)$		
				 & $\left(\begin{array}{ccc}  2 & 2  & 0  \\ 2  & 2  & 0  \\ 2  & 2  & -16  \end{array}\right)$  \\
& $a_{22i}^j$		& $\left(\begin{array}{ccc}  8 & 0  & -24  \\ 0  & 8  & -24  \\ 0  & 0  & -16  \end{array}\right)$	& 
				$\left(\begin{array}{ccc}  0 & 0  & 0  \\ 0  & 0  & 0  \\ 0  & 0  & 0  \end{array}\right)$	& $\left(\begin{array}{ccc}  0 & 2  & 0  \\ 0  & 2  & 0  \\ 0  & 2  & 0  \end{array}\right)$\\
& $a_{2i3}$		& $(8,8,8)$	& $(0,0,0)$ 	& $(0,0,0)$		\\
\hline

P & $a_{3i0}$		& $(0,0,0)$	& $(0,0,4)$ 	& $(0,0,16)$			\\	
& $a_{31i}^j$		& $\left(\begin{array}{ccc}  0 & 0  & 0  \\ 0  & 0  & 0  \\ 0  & 0  & 0  \end{array}\right)$		& 
				$\left(\begin{array}{ccc}  4 & 0  & 4  \\ 0  & 4  & 4  \\ 0  & 0  & 4  \end{array}\right)$& $\left(\begin{array}{ccc}  2 & 2  & 0  \\ 2  & 2  & 0  \\ 2  & 2  & -16  \end{array}\right)$		  \\
& $a_{32i}^j$		& $\left(\begin{array}{ccc}  0 & 0  & 0  \\ 0  & 0  & 0  \\ 0  & 0  & 0  \end{array}\right)$		& 
				$\left(\begin{array}{ccc}  0 & 12  & 4  \\ 12  & 0  & 4  \\ 12  & 12  & 16  \end{array}\right)$& $\left(\begin{array}{ccc}  0 & 2  & 0  \\ 0  & 2  & 0  \\ 0  & 2  & 0  \end{array}\right)$		 \\
& $a_{3i3}$		& $(0,0,0)$	& $(-12,-12,8)$		& $(0,0,0)$			\\
\hline\hline
$V_{min}$ &		&$-1.1244$	&$-647.35$	&$-0.35413$	\\
\hline
\end{tabular}
\end{center}
\end{table}%

\begin{table}[htdp]
\caption{\label{tab:afluxes} Flux configuration for the stable AdS cases (anisotropic).}
\begin{center}
\begin{tabular}{c|c|cc}
\hline
Sector &Flux 		& A-A-Q		& A-A-PQ		\\
\hline
RR & $a_{000}$	&$0$			&$0$			\\
& $a_{01i}$		&$(0,50,8)$	&$(0,0,0)$		\\
& $a_{02i}$		&$(58,0,14)$	&$(0,0,6)$		\\
& $a_{033}$		&$12$		&$4$			\\
\hline
NSNS & $a_{100}$	&$8$			&$4$			\\	
& $a_{11i}$		&$(40,18,8)$	&$(12,60,16)$	\\
& $a_{12i}$		&$(18,40,38)$	&$(0,12,60)$	\\
& $a_{133}$		&$38$		&$0$	\\
\hline
Q & $a_{2i0}$		&$(-8,0,0)$	&$(0,0,0)$		\\	
& $a_{21i}^j$		& $\left(\begin{array}{ccc}  0 & 0  & 6  \\ 0  & -8  & 4  \\ 0  & 8  & 0  \end{array}\right)$ 
				& $\left(\begin{array}{ccc}  8 & -24  & 0  \\ 0  & 0  & 0  \\ 0  & 0  & 0  \end{array}\right)$\\
& $a_{22i}^j$		& $\left(\begin{array}{ccc}  0 & 8  & 4  \\ 0  & 0  & 6  \\ 0  & 2  & -6  \end{array}\right)$ 
				& $\left(\begin{array}{ccc}  0 & 0  & 0  \\ 6  & 18  & 0  \\ -8  & 24  & 0  \end{array}\right)$\\
& $a_{2i3}$		&$(0,2,6)$		&$(0,0,0)$		\\
\hline

P & $a_{3i0}$		&$(0,0,0)$		&$(0,0,0)$		\\	
& $a_{31i}^j$		&$\left(\begin{array}{ccc}  0 & 0  & 0  \\ 0  & 0  & 0  \\ 0  & 0  & 0  \end{array}\right)$ 
				&$\left(\begin{array}{ccc}  60 & 0  & 0  \\ 0  & -36  & 0  \\ 0  & 0  & 0  \end{array}\right)$\\
& $a_{32i}^j$		&$\left(\begin{array}{ccc}  0 & 0  & 0  \\ 0  & 0  & 0  \\ 0  & 0  & 0  \end{array}\right)$
				&$\left(\begin{array}{ccc}  -60 & 36  & 0  \\ 0  & 0  & 0  \\ 24  & 40  & 52  \end{array}\right)$\\
& $a_{3i3}$		&$(0,0,0)$		&$(24,40,-52)$	\\
\hline\hline
 $V_{min}$& 		&$-0.00310$	&$-28.257$	\\
\hline
\end{tabular}
\end{center}
\end{table}%

With the purpose to compare more date among the different compactification scenarios, it is illustrative to show the mass of moduli, the F-terms and the mass of gravitinos. The corresponding mass for all the 6 real stabilized moduli after diagonalization of the mass matrix (see \cite{Damian:2013dq}) for details) given by
\begin{equation}
M^2_{ij}=D_i\partial_j\mathscr{V}+\frac{2}{3}V_0{\mathscr K}_{ij},
\label{massnd}
\end{equation}
 where $D_i$ stands for the K\"ahler derivate, are  shown in Table \ref{tab:massesads} for the complete set of 11 different vacua, while the F-terms 
\begin{equation}
<F^i>=<e^{\mathscr{K}}{\cal D}_{\bar{i}} \mathscr{W}\mathscr{K}^{i\bar{i}}>,
\end{equation}
and the mass of gravitinos given by $m_{3/2}=e^{\mathscr{K}/2}W$ are presented in Table \ref{tab:gravAdS_m}. We observe that the energy vev and the moduli mass seem to be independent  either of the flux configuration and/or of the type of torus compactification (isotropic, semi-isotropic or anisotropic). Also it is important to remark that the absence of anisotropic solutions might be a consequence of the method we are using.\\

\begin{table}[htdp]
\caption{\label{tab:massesads}Moduli squared masses for AdS vacua.}
\begin{center}
\begin{tabular}{c|cccccc}
\hline
Isotropic tori&&&&&&\\

Moduli 		& A-I-P1			& A-I-P2			& A-I-P3 		&A-I-QP1 			& A-I-QP2	&A-I-QP3 	\\
\hline
$\phi_1$				& $1.880{\times}10^5$		& $2.332{\times}10^4$		& $19191$	& $2.766{\times}10^6$	& $3.8103{\times}10^6$		& $1425.3$		\\
$\psi_1$				& $1.202{\times}10^5$		& $2.177{\times}10^4$		& $18913$	& $2.7484{\times}10^6$	& $3.8085{\times}10^6$		& $1100.8$		\\
$\phi_4$				& $4.656{\times}10^4$		& $1.065{\times}10^4$		& $6198.4$	& $3379.7$			& $315.45$				& $371.90$		\\
$\psi_4$				& $3.321{\times}10^4$		& $3.889{\times}10^3$		& $2496.3$	& $1753.8$				& $185.52$				& $151.62$	\\
$\phi_5$				& $6.347{\times}10^3$		& $46.450$				& $20.054$	& $13.042$				& $0.5605$				& $40.286$		\\
$\psi_5$				& $4.983{\times}10^3$		& $25.422$				& $12.727$	& $12.748$				& $0.55676$				& $39.443$		\\
\hline
Semi-isotropi tori&&&&&&\\
Moduli   &  A-S-Q &A-S-P& A-S-QP&&&\\
\hline
$\phi_{1,2}$	& $94.354$		& $33884$		& $5.6582$	\\
$\psi_{1,2}$	& $36.457$		& $26047$		& $4.2746$	\\
$\phi_3$		& $14.518$		& $9341.5$		& $1.4939$\\
$\psi_3$		& $3.3614$		& $4970.9$		& $0.6848$\\
$\phi_4$		& $0.9472$		& $780.0$			& $0.4497$\\
$\psi_4$		& $0.8651$		& $344.1$			& $0.3645$\\
$\phi_{5,6}$	& $0.6345$		& $1072.5$		& $0.2575$\\
$\psi_{5,6}$	& $0.7627$		& $365.14$                & $0.1582$\\
$\phi_7$       	& $0.7411$                & $464.24$		& $0.2180$\\
$\psi_7$	         & $0.7515$		& $452.55$		& $0.2395$	\\
\hline
Anisotropic tori\\
Moduli 		&A-A-Q				& A-A-QP \\
\hline
$\phi_1$		& $1.0068$			& $4081.9$			\\
$\psi_1$		& $1.0068$			& $3936.5$			\\
$\phi_2$		& $2.1544\times 10^{-3}$	& $271.39$			\\
$\psi_2$		& $2.1405\times 10^{-3}$	& $131.35$			\\
$\phi_3$		& $2.1246\times 10^{-3}$	& $64.055$			\\
$\psi_3$		& $2.0890\times 10^{-3}$	& $44.371$			\\
$\phi_4$    	& $2.0735\times 10^{-3}$	& $6.902$				\\
$\psi_4$		& $2.0714\times 10^{-3}$	& $26.353$			\\
$\phi_5$		& $2.0673\times 10^{-3}$	& $12.650$			\\
$\psi_5$		& $2.0698\times 10^{-3}$	& $23.154$			\\
$\phi_2$		& $2.0692\times 10^{-3}$	& $19.936$			\\
$\psi_2$	   	& $2.0686\times 10^{-3}$	& $18.087$			\\
$\phi_7$		& $2.0787\times 10^{-3}$	& $19.198$			\\
$\psi_7$		& $2.0687\times 10^{-3}$	& $18.877$			\\
\hline
\end{tabular}
\end{center}
\end{table}

\begin{table}[htdp]
\caption{\label{tab:gravAdS_m} Gravitino mass and F-term for each AdS case.}
\begin{center}
\begin{tabular}{ccc}
\hline
Case		& $m_{3/2}^2$		& $F-$term\\
\hline
A-I-P1		& $29.09$			& $8.5360$			\\
A-I-P2		& $9.14$			& $193.58$			\\
A-I-P3		& $8.30$			& $343.42$			\\
A-I-QP1		& $48.19$			& $1.95\times{10^5}$	\\
A-I-QP2		& $15.89$			& $2.53\times{10^6}$	\\
A-I-QP3		& $12.72$			& $11248$			\\
A-S-Q		& $4.10$			& $76484$			\\
A-S-P		& $14.68$			& $0$				\\
A-S-QP		& $0.69$			& $7629.1$			\\
A-A-Q		& $0.09$			& $3.10\times{10^5}$	\\
A-A-QP		& $11.63$			& $1.00\times{10^5}$	\\

\hline
\end{tabular}
\end{center}
\end{table}

\subsection{De-Sitter vacua}
We find 7 explicit flux configurations fulfilling the tadpole and Bianchi constraints with a positive-valued minimum in the absence of sources.
There are two solutions which correspond to compactifications on isotropic torus with only $P$-fluxes (dS-I-P) and with both non-geometric fluxes $(Q,P)$ (dS-I-QP). We have also found 2 vacua related to compactifications on semi-isotropic torus with  $Q$  (dS-S-Q) and with both $Q$ and $P$ -fluxes (dS-S-QP). In a compactification on an anisotropic torus, we report also two vacua, concerning a compactification threaded only with  $Q$-fluxes (dS-A-Q) and with $P$-fluxes (dS-S-P) respectively. Our results are shown in the Tables \ref{tab: ds}  and \ref{tab:Sfluxes}.\\
\begin{table}[htdp]
\caption{\label{tab: ds} Flux configuration for dS vacua.}
\begin{center}
\begin{tabular}{c|c|cccc}
\hline
		&Flux  			& dS-I-P		&dS-I-QP		\\
\hline
RR		&$a_{00}^0$		& $20$		&$4$	 		\\
      		&$a_{01}^i$		& $36$		&$8$	 		\\
      		&$a_{02}^{ij}$		& $44$		&$20$   		\\
      		&$a_{03}^{ijk}$		& $48$		&$53$   		\\
\hline
NS-NS	&$a_{10}^0$		& $50$		&$-4$		\\	
		&$a_{11}^i$		& $42$		&$0$			\\
		&$a_{12}^{ij}$		& $34$		&$0$			\\
		&$a_{13}^{ijk}$		& $28$		&$-50$		\\
\hline
$Q$		&$a_{20}^{m0}$	&$0$ 		& $0$		\\	
		&$a_{21}^{mi}$		&$(0,0,0)$		& $(-12,-12,0)$	\\
		&$a_{22}^{mij}$	&$(0,0,0)$		& $(0,0,0)$	\\
		&$a_{23}^{mijk}$	&$0$			& $60$		\\
\hline
$P$		&$a_{30}^{m0}$	& $-40$		&$0$			\\	
		&$a_{31}^{mi}$		& $(20,20,-40)$&$(0,0,0)$		\\
		&$a_{32i}^{mij}$	& $(20,20,-10)$	&$(0,0,30)$ 	\\
		&$a_{33}^{mijk}$	& $10$		&$0$			\\
\hline
\hline

		&$V_{min}$ 		& $1005.4$	& $811.81$	\\
\hline
\end{tabular}
\end{center}
\end{table}

\begin{table}[htdp]
\caption{\label{tab:Sfluxes} Flux configuration for the stable dS cases.}
\begin{center}
\begin{tabular}{ccccccc}
\hline
Sector & Flux 			& dS-S-Q 		& dS-S-QP 		& dS-A-Q		&dS-A-P  			\\
\hline
RR & $a_{00}$			& $0$		& $0$ 		& $0$		& $0$		 	\\
& $a_{01}$			& $(0,0,30)$	& $(0,0,2)$ 	& $(0,8,8)$	& $(0,0,0)$	 	\\
& $a_{02}$			& $(-56,-56,0)$	& $(0,0,0)$ 	& $(24,0,16)$	& $(20,58,24)$	 	\\
& $a_{03}$			& $34$		& $0$ 		& $8$		& $54$			\\
\hline
NSNS & $a_{10}$		& $-34$		& $8$ 		& $2$		& $8$			\\	
& $a_{11}$			& $(2,2,2)$	& $(16,16,14)$ 	& $(26,54,2)$	& $(0,8,0)$		\\
& $a_{12}$			& $(30,30,30)$	&$(-16,-16,16)$	&$(54,26,50)$	& $(16,52,32)$		\\
& $a_{13}$			& $-62$		& $16$ 		& $50$ 		& $56$			\\
\hline
Q & $a_{20}$			& $(0,0,26)$	& $(0,0,58)$ 	&$(0,-22,0)$	& $(0,0,0)$	  	\\
& $a_{31}^j$			& $\left(\begin{array}{ccc}  8 & 8  & 2  \\ 8  & 8  & 2  \\ 8  & 8  & -50  \end{array}\right)$
					& $\left(\begin{array}{ccc}  2 & 2  & 0  \\ 2  & 2  & 0  \\ 2  & 2  & -58  \end{array}\right)$ 
					& $\left(\begin{array}{ccc}  0 & 0  & 4  \\ 0  & -44  & 18  \\ 0  & 22  & -36  \end{array}\right)$ 
					& $\left(\begin{array}{ccc}  0 & 0  & 0  \\ 0  & 0  & 0  \\ 0  & 0  & 0  \end{array}\right)$	\\
& $a_{32}^j$			& $\left(\begin{array}{ccc}  16 & 0  & -26  \\ 0  & 16  & -26  \\ 0  & 0  & -22  \end{array}\right)$ 
					& $\left(\begin{array}{ccc}  2 & 0  & 0  \\ 0  & 2  & 0  \\ 0  & 0  & 0  \end{array}\right)$ 
					& $\left(\begin{array}{ccc}  22 & 18  & 4  \\ 44  & 0  & 18  \\ 0  & 4  & -36  \end{array}\right)$ 
					& $\left(\begin{array}{ccc}  0 & 0  & 0  \\ 0  & 0  & 0  \\ 0  & 0  & 0  \end{array}\right)$	\\
& $a_{33}$			& $(8,8,2)$	& $(0,0,0)$ 	& $(0,22,36)$	& $(0,0,0)$	 	\\

\hline
P & $a_{30}$			& $(0,0,0)$	& $(0,0,58)$ 	& $(0,0,0)$	& $(0,0,0)$		\\	
& $a_{31}^j$			& $\left(\begin{array}{ccc}  0 & 0  & 0  \\ 0  & 0  & 0  \\ 0  & 0  & 0  \end{array}\right)$	
					& $\left(\begin{array}{ccc}  2 & 2  & 0  \\ 2  & 2  & 0  \\ 2  & 2  & -58  \end{array}\right)$ 
					& $\left(\begin{array}{ccc}  0 & 0  & 0  \\ 0  & 0  & 0  \\ 0  & 0  & 0  \end{array}\right)$ 
					& $\left(\begin{array}{ccc}  0 & 0  & 0  \\ 0  & 0  & 0  \\ 0  & 0  & 0  \end{array}\right)$	\\
& $a_{32}^j$			& $\left(\begin{array}{ccc}  0 & 0  & 0  \\ 0  & 0  & 0  \\ 0  & 0  & 0  \end{array}\right)$	 
					& $\left(\begin{array}{ccc}  2 & 0  & 0  \\ 0  & 2  & 0  \\ 0  & 0  & 0  \end{array}\right)$ 
					& $\left(\begin{array}{ccc}  0 & 0  & 0  \\ 0  & 0  & 0  \\ 0  & 0  & 0  \end{array}\right)$ 
					& $\left(\begin{array}{ccc}  44 & -8  & 60  \\ 0  & 0  & 0  \\ 0  & 0  & 0  \end{array}\right)$\\
& $a_{33}$			& $(0,0,0)$	& $(0,0,0)$ 	& $(0,0,0)$		& $(44,0,0)$			 \\
\hline
& $V_{min}$ 			& $6.5858$	& $0.00563$	& $4.14554$		&$8.1689{\times}10^{-4}$		\\
\hline
\end{tabular}
\end{center}
\end{table}%

Since we have not imposed extra conditions on the fluxes, we have assumed that SUSY is broken through all moduli. However it is observed that  the F-terms related to the complex structure moduli have the higher value, suggesting that SUSY is broken mainly by this complex field and therefore, establishing that the sgoldstino points towards this direction \cite{Borghese:2012yu, Damian:2013dq}. \\

As in the AdS cases, 
 the F-terms and the gravitino mass present high values for phenomenological purposes.  Concerning the F-terms, for all cases they
are of order $10^3$ in Planck units, while the gravitinos mass for the 4 cases are of order of $10^1-10^2$ as shown in Table  \ref{tab:grav_m}.
\begin{table}[htdp]
\caption{\label{tab:grav_m} Gravitino mass and F-term for each dS case.}
\begin{center}
\begin{tabular}{ccc}
\hline
Case		& $m_{3/2}^2$		& $F-$term\\
\hline
dS-I-P		& $194.32$		& $29455$		\\
dS-I-QP		& $23.36$			& $4632.9$		\\
dS-S-Q		& $4.6538$		& $20271$			\\
dS-S-QP		& $1.1082$		& $1142.5$			\\
dS-A-Q		& $4.4625$		& $1498.8$			\\
dS-A-P		& $13.135$		& $8267.7$				\\
\hline
\end{tabular}
\end{center}
\end{table}
These values seems to be also uncorrelated to the specific features of the model. There is however one issue we want to remark: all stable cases are within the BF limit, meaning that the moduli squared mass is positive due to the curvature contribution of the dS space-time to the second derivative of the potential \cite{Abbott:1981ff}, i.e., from Eq.(\ref{massnd})
\begin{equation}
D_i\partial_j\mathscr{V}<0,
\end{equation}
for all our solutions. The mass of all stabilized moduli are given in Table \ref{tab:massds}, where the curvature contribution to the mass 
has been already considered as well as the required diagonalization of the mass matrix. In consequence, we can notice  the following:

\begin{itemize}
\item
In all cases a mass hierarchy on the moduli is present. Specifically  it seems that  most of the complex structure moduli are heavier than the rest. We also notice that for the isotropic and anisotropic cases, the complex structure moduli posses an  upper-Hubble mass, implying that they cannot drive inflation in the small-field regime. For the isotropic cases,   there are sub-Hubble masses  corresponding to the K\"ahler moduli masses. This suggests the possibility that in case of having suitable conditions for small-field inflation, it would be driven by the K\"ahler moduli.
For the anisotropic cases we see that for both cases the sub-Hubble masses correspond to the K\"ahler moduli and to the imaginary part of the axio-dilaton. In this sense we can conclude that at least for these two cases, the chances to identify a possible slow-roll inflaton in the small-field regime increases. Notice that the value of the  sub-Hubble masses of the K\"ahler  fields come mainly from the curvature term in the mass formula \ref{massnd}.
\item
For the isotropic and anisotropic tori,  the heaviest moduli correspond to the complex structure moduli, which also break SUSY. Therefore,  in such  models,  the sgoldstino is almost orthogonal to possible inflationary directions.
\item
For the semi-isotropic torus,  we have found different features with respect to the above two cases: For the (dS-S-Q) case, there are moduli fields in all sectors, K\"ahler, dilaton and complex structure, which have sub-Hubble mass. In suitable conditions, all of them are candidate for driven small-field inflation. Also we notice that for the (dS-S-P) case, all moduli are heavier than the Hubble scale, implying that in such model,  inflation is not present.
\item
For the anisotropic case, we report the existence of 2 stable vacua
\end{itemize}

\begin{table}[htdp]
\caption{\label{tab:massds} Moduli squared masses for dS vacua.}
\begin{center}
\begin{tabular}{c|ccccc|}
\hline
Isotropic tori	&$H_0=1005.4$		& $H_0=811.81$		& $H_0=6.5858$\\
Moduli 		& dS-I-P 			         & dS-I-QP				& dS-S-Q		\\
\hline
$\phi_1$		& $1.093{\times}10^6$	& $8.7741{\times}10^4$	& $112.53$	\\
$\psi_1$		& $9.133{\times}10^5$	& $2.8076{\times}10^4$	& $6.6042$	\\
$\phi_2$		&					&					& $$			\\
$\psi_2$		&					&					& $$			\\
$\phi_3$		&					&					& $80.243$	\\
$\psi_3$		&					&					& $5.3844$	\\	
\hline
$\phi_4$		& $2.132{\times}10^4$	& $1.9624\times 10^4$	& $0.95858$	\\
$\psi_4$		& $1.548{\times}10^4$	& $1.7847\times 10^4$	& $5.1772$	\\
\hline
$\phi_5$		& $5.588{\times}10^2$	& $6.6473\times 10^2$	& $0.1953$	\\
$\psi_5$		& $7.186 {\times}10^2$	& $5.9443\times 10^2$	& $4.5082$	\\
$\phi_6$		&					&					& $$			\\
$\psi_6$		&					&					& $$			\\	
$\phi_7$		&					&					& $8.4962$	\\	
$\psi_7$		&					&					& $4.2906$	\\	
\hline
\hline
Semi and anisotropic tori&$H_0=0.00563$	&$H_0=4.1455$	&$H_0=8.1689{\times}10^{-4}$		\\
Moduli 		& dS-S-QP			&dS-A-Q		&dS-A-P				\\
\hline
$\phi_1$		& $183.90$		&$717.10$	& $79.6350$			\\
$\psi_1$		& $173.22$		&$597.23$	& $60.0020$			\\
$\phi_2$		& $$				&$223.90$	& $25.9790$			\\
$\psi_2$		& $$				&$126.55$	& $6.42500$			\\
$\phi_3$		& $15.699$		&$29.663$	& $7.803\times 10^{-2}$	\\
$\psi_3$		& $10.812$		&$15.838$	& $1.019\times 10^{-2}$	\\
\hline
$\phi_4$		&$6.7188$		&$5.993$		& $1.24\times 10^{-3}$	\\
$\psi_4$		& $3.5458$		&$10.536$	& $7.3\times 10^{-4}$	\\
\hline
$\phi_5$		& $0.74141$		&$2.6369$	& $5.77\times 10^{-3}$	\\
$\psi_5$		& $0.14348$		&$6.0460$	& $5.39\times 10^{-3}$	\\
$\phi_6$		& $$				&$4.3800$	& $5.48\times 10^{-3}$	\\
$\psi_6$		& $$				&$2.4681$	& $5.4\times 10^{-4}$	\\
$\phi_7$		& $0.10034$		&$3.2509$	& $5.4\times 10^{-4}$	\\
$\psi_7$		& $0.05992$		&$2.9891$	& $5.4\times 10^{-4}$	\\
\hline

\end{tabular}
\end{center}
\end{table}

It is worth to compare the above results with the reported solutions in \cite{Damian:2013dq}, corresponding to dS vacua constructed from an isotropic torus in the presence of $Q$-fluxes. The (squared) masses of gravitinos here reported are of the order $10^1-10^{3}$ in Planck units, while those found in \cite{Damian:2013dq} are of order $10^{-4}$. Therefore, for dS vacua, the most attractive model for phenomenological reasons so far obtained, are those constructed by compactification on isotropic torus with $Q$-fluxes and vanishing $P$-fluxes.

\section{Final Comments}
In this work we report the existence of  stable AdS and dS vacua in the context of  type IIB string compactification on isotropic, semi-isotropic and anisotropic tori threaded with RR, NS-NS and S-dual $(Q,P)$ non-geometric fluxes in the presence of orientifold 3-planes. The case of $P=0$ on an isotropic torus was studied  in \cite{Damian:2013dq}.
By implementing a genetic algorithm we have been able to find 11 AdS and 7 dS stable vacua.  Although our algorithm does not allows us to search the entire flux configuration landscape, we notice some generalities of the reported solutions:
\begin{enumerate}
\item
By looking for stable vacua on anisotropic tours compactification, we report 2 {\it fully} stable (mod torus factorization) dS vacua in the BF limit, meaning that their tachyonic behavior is erased by the space-time curvature contribution.
\item
In \cite{Dibitetto:2011gm} the authors report the existence of stable solutions in the ${\cal N}=1$ subset of scalars in the supergravity effective theory  but unstable  with respect to the full  ${\cal N}=4$ theory. It would be very interesting wether the solutions here presented maintain their stable status once they uplift to ${\cal N}$=4 supergravity by considering the whole set of moduli.
\item
The complex structure masses are always larger than the axio-dilaton and K\"ahler moduli masses for both, the isotropic and anisotropic cases. The F-term related to the complex structure moduli is always larger than the rest, suggesting that the sgoldstino is directed along the complex structure derivative and since the masses are always larger than the Hubble scale, the complex structure is discarded as a possible inflaton in the small-field regime.
\item
The order of magnitude of the energy vev's, gravitino masses and F-terms seems to be independent of the isotropy or anisotropy of the 6-dimensional torus and of the presence of both non-geometric fluxes $(Q,P)$ or just one kind of  them $Q$ or $P$. However we notice that the number moduli masses which are below the Hubble scale increases in the anisotropic case compared with the isotropic case. Therefore, at least for the cases we report, we observe that the size of sub-Hubble moduli increases in the anisotropic compactifications suggesting  that in such scenarios the probability to find candidates for small-field inflation is also bigger, although deeper studies are needed.
\end{enumerate}

The models here presented corresponds only to a minuscule portion of the whole field configuration space. However we believe according to the form of the superpotential which clearly establishes a different role for the complex structure with respect to the other moduli, that the above features are rather generic. Clearly it is necessary to explore in more detail the space of solutions. It might be possible that  corrections to the K\"ahler potential and/or incorporation of D-branes would change the values of masses and SUSY breaking scales to more realistic scenarios.

\begin{center}
{\bf Acknowledgements}
\end{center}
We thank Alfonso Guarino, Ralph Blumenhagen and Ivonne Zavala for useful comments and suggestions and to Gustavo Niz, Miguel Sabido and Saul Ramos-Sanchez for many interesting discussions. D. A is supported by a doctoral CONACyT grant. O.L.-B. is partially supported by CONACyT under contract No. 132166 and by PROMEP.

\appendix
\section{Genetic code}
The following code shows the methodology employed to minimize the scalar potential for the anisotropic case. The remaining cases are taken as particular cases of this algorithm. The source code assumes that a scalar potential is stored on a handled function called \verb V_anisotropic,  whose entries correspond to the fluxes which satisfy both tadpole cancellation conditions and the Bianchi identities. The final points of interest given by the genetic algorithm are stored on a file called \verb sol_anisotropic.m  and \verb fluxes_anisotropic.m.  Once these variables are stored, the function "ToMathematica" translates all data into a Mathematica language. Finally, these solutions are further improved in mathematica by using the standard Newton method with an accuracy of 1000 decimals\footnote{We thank Saul Ramos-Sanchez for suggesting us this improvement in our method.} (the source code is not shown).\\
\begin{mylisting}
\begin{verbatim}
clear all;
clc
\%up and lb contains the boundaries on the moduli space explored by the ga
k=3;											\%Family solution
lb=[.1,.0001,-500,.0001,-500,0.0001,-500,0.0001,...
			-500,0.001,-500,0.001,-500,0.001];		\%Lower bound
up=[1e10,1e10,1e10,1e10,1e10,1e10,1e10,1e10,...
			1e10,1e10,1e10,1e10,1e10,1e10]; 		\%Upper bound
nvar=size(lb,2); 								\%Number of variables
max=2000;									\%Maximum number of iterations
time_ga_parallel=0;
sol=zeros(max,15);
fluxes=zeros(max,64);



fprintf('\n Minimization process begins\n');
for i=1:max
        [a0,a11,a12,a13,a21,a22,a23,a3,...
            b0,b11,b12,b13,b21,b22,b23,b3,...
            c01,c02,c03,c111,c112,c113,c121,c122,c123,c131,c132,c133,...
            c211,c212,c213,c221,c222,c223,c231,c232,c233,c31,c32,c33,...
            d01,d02,d03,d111,d112,d113,d121,d122,d123,d131,d132,d133,...
            d211,d212,d213,d221,d222,d223,d231,d232,...
            d233,d31,d32,d33=constraint_solution(3,32);
        fluxes(i,:)=[a0,a11,a12,a13,a21,a22,a23,...
            a3,b0,b11,b12,b13,b21,b22,b23,b3,...
            c01,c02,c03,c111,c112,c113,c121,c122,c123,c131,c132,c133,...
            c211,c212,c213,c221,c222,c223,c231,c232,c233,c31,c32,c33,...
            d01,d02,d03,d111,d112,d113,d121,d122,d123,d131,d132,d133,...
            d211,d212,d213,d221,d222,d223,d231,d232,d233,d31,d32,d33];
       

       %Here I call the function V_anisotropic.m which is a handle function
      
        V=V_anisotropicC(a0,a11,a12,a13,a21,a22,a23,...
            a3,b0,b11,b12,b13,b21,b22,b23,b3,...
            c01,c02,c03,c111,c112,c113,c121,c122,c123,c131,c132,c133,...
            c211,c212,c213,c221,c222,c223,c231,c232,c233,c31,c32,c33,...
            d01,d02,d03,d111,d112,d113,d121,d122,d123,d131,d132,d133,...
            d211,d212,d213,d221,d222,d223,d231,d232,d233,d31,d32,d33);
        options = gaoptimset('Generations',500,'TolCon',1e-6,'TolFun',1e-10,...
        			'PopulationSize',200,'MutationFcn',@mutationadaptfeasible,...
			'StallGenLimit',20,'Display','off','Vectorized','on');
        startTime = tic;
        [x,fval,flag]=ga(V,nvar,[],[],[],[],lb,[],[],options);
        time_ga_parallel = toc(startTime);
        fprintf('Iter. \%g takes \%g minutes.',i,time_ga_parallel/60.);
        
        
        fprintf('\n SolFluxes={a0->\%d,a11->\%d,a12->\%d,a13->\%d,a21->\%d,a22->\%d,...
        a23->\%d,a3->\%d,b0->\%d,b11->\%d,b12->\%d,b13->\%d,b21->\%d,b22->\%d,...
        b23->\%d,b3->\%d,c01->\%d,c02->\%d,c03->\%d,c111->\%d,c112->\%d,c113->\%d,...
        c121->\%d,c122->\%d,c123->\%d,c131->\%d,c132->\%d,c133->\%d,c211->\%d,...
        c212->\%d,c213->\%d,c221->\%d,c222->\%d,c223->\%d,c231->\%d,c232->\%d,...
        c233->\%d,c31->\%d,c32->\%d,c33->\%d,d01->\%d,d02->\%d,d03->\%d,d111->\%d,...
        d112->\%d,d113->\%d,d121->\%d,d122->\%d,d123->\%d,d131->\%d,d132->\%d,...
        d133->\%d,d211->\%d,d212->\%d,d213->\%d,d221->\%d,d222->\%d,...
        d223->\%d,d231->\%d,d232->\%d,d233->\%d,d31->\%d,d32->\%d,d33->\%d};',fluxes(i,:));
        fprintf('\ntestsol={taur1->\%0.25f,taui1->\%0.25f,taur2->\%0.25f,taui2->\%0.25f,...
        taur3->\%0.25f,taui3->\%0.25f,Sr->\%0.25f,Si->\%0.25f,Ur1->\%0.25f,Ui1->\%0.25f,...
        Ur2->\%0.25f,Ui2->\%0.25f,Ur3->\%0.25f,Ui3->\%0.25f};\n',x);
        sol(i,:)=[x,fval];
        
end
figure(1);
loglog(1./sol(:,8),-sol(:,15),' x ');
xlabel('g_s');
ylabel('V_0');
title('Negative Lambda');
figure(2);
loglog(1./sol(:,8),sol(:,15),' x ');
xlabel('g_s');
ylabel('V_0');
title('Positive Lambda');

% Here I write the final solution *************************
save('sol_anisotropic','sol');                             
save('fluxes_anisotropic','fluxes');                            
%***********************************************************
ToMathematica

\end{verbatim}
\end{mylisting}

\section{AdS vacua from an isotropic torus with $Q$-fluxes.}

For completeness we show in Table \ref{tab:fluxesAdSIQ} the explicit flux configurations for stable AdS vacua obtained in our previous work \cite{Damian:2013dq}. They correspond to compactifications of Type IIB on an isotropic torus threaded with $Q$-fluxes ($P$-fluxes are absent) in the presence of orientifold 3-planes. All of them are non-SUSY and SUSY is broken through all moduli.

\begin{table}[htdp]
\caption{\label{tab:fluxesAdSIQ} Flux configuration for AdS vacua. All cases correspond to compactification on an isotropic T-dual torus. The value for the scalar potential at its minimum is given in mass Planck units.}
\begin{center}
\begin{tabular}{c|c|cccc}
\hline
&Flux 			& A-I-Q1		  & A-I-Q2 		& A-I-Q3	& A-I-Q4		\\
\hline
RR&$a_{00}^0$	& $2$		& $2$		& $2$	& $2$		\\
&$a_{01}$		& $18$ 		& $16$		& $20$	& $16$ 		\\
&$a_{02}$		& $2$ 		& $14$		& $18$	& $8$		\\	
&$a_{03}$		& $20$		& $24$		& $38$	& $34$		\\

\hline
NS-NS&$a_{13}$		& $16$		& $16$		& $16$	& $16$		\\
\hline
Q&$a_{22}^j$		&$(0,0,2)$		&$(0,0,2)$		&$(0,0,4)$	& $(0,0,4)$	\\
&$a_{23}$		&$18$		&$16$		&$40$	& $32$		\\
\hline
\hline
&$V_{min}$ 		& $-0.0290$	& $-0.0259$	& $-0.0222$&$-6.775{\times}10^{-7}$	\\
\hline
\end{tabular}
\end{center}
\end{table}%

The corresponding moduli masses for each case are shown in Table \ref{tab:massesadsQ}, while the gravitino masses and F-terms are shown in Table \ref{tab:gravAdS_m}

\begin{table}[htdp]
\caption{\label{tab:massesadsQ}Moduli squared masses for AdS vacua.}
\begin{center}
\begin{tabular}{c|cccccc}
\hline
Isotropic tori&&&&&&\\

Moduli 	& A-I-Q1		& A-I-Q2		& A-I-Q3 		&A-I-Q4 			 		\\
\hline
$\phi_1$	& $237.926$	& $2.100$		& $143.972$	& $1.188{\times}10^{-4}$		\\
$\psi_1$	& $3.251$		& $0.108$		& $2.609$		& $1.110{\times}10^{-5}$		\\
$\phi_4$	& $191.305$	& $0.997$		& $5.007$		& $4.517{\times}10^{-7}$		\\
$\psi_4$	& $1.102$		& $0.0259$	& $0.0902$	& $4.527{\times}10^{-7}$		\\
$\phi_5$	& $0.0197$	& $0.0017$	& $0.0147$	& $4.517{\times}10^{-7}$		\\
$\psi_5$	& $0.0198$	& $0.0017$	& $0.0147$	& $4.521{\times}10^{-7}$		\\
\hline
\end{tabular}
\end{center}
\end{table}

\begin{table}
\caption{\label{tab:gravAdS_m2} Gravitino mass and F-term for each AdS case.}
\begin{center}
\begin{tabular}{ccc}
\hline
Case		& $m_{3/2}^2$		& $F-$term\\
\hline
A-I-Q1	& $0.1511$		& $666.034$			\\
A-I-Q2	& $0.0470$		& $708.784$			\\
A-I-Q3	& $0.1542$		& $1358.15$			\\
A-I-Q4	& $0.0008$		& $1087.46$	\\
\hline
\end{tabular}
\end{center}
\end{table}

\bibliography{anisotropic}

\providecommand{\bysame}{\leavevmode\hbox to3em{\hrulefill}\thinspace}
\begin{thebibliography}{10}

\bibitem{Kachru:2003aw}
S.~Kachru, R.~Kallosh, A.~D. Linde, and S.~P. Trivedi, \emph{{De Sitter vacua
  in string theory}}, Phys.Rev. \textbf{D68} (2003), 046005,
  \texttt{hep-th/0301240}.

\bibitem{Kallosh:2007ig}
R.~Kallosh, \emph{{On inflation in string theory}}, Lect.Notes Phys.
  \textbf{738} (2008), 119--156,  \texttt{hep-th/0702059}.

\bibitem{Andriot:2013txa}
D.~Andriot, \emph{{Non-geometric fluxes versus (non)-geometry}},  (2013),
  \texttt{1303.0251}.

\bibitem{Danielsson:2009ff}
U.~H. Danielsson, S.~S. Haque, G.~Shiu, and T.~Van~Riet, \emph{{Towards
  Classical de Sitter Solutions in String Theory}}, JHEP \textbf{0909} (2009),
  114,  \texttt{0907.2041}.

\bibitem{Danielsson:2010bc}
U.~H. Danielsson, P.~Koerber, and T.~Van~Riet, \emph{{Universal de Sitter
  solutions at tree-level}}, JHEP \textbf{1005} (2010), 090,
  \texttt{1003.3590}.

\bibitem{Chen:2011ac}
X.~Chen, G.~Shiu, Y.~Sumitomo, and S.~Tye, \emph{{A Global View on The Search
  for de-Sitter Vacua in (type IIA) String Theory}},  (2011),
  \texttt{1112.3338}, 22 pages, 5 figures/ v2, v3: arguments improved,
  references added.

\bibitem{Shiu:2011zt}
G.~Shiu and Y.~Sumitomo, \emph{{Stability Constraints on Classical de Sitter
  Vacua}}, JHEP \textbf{1109} (2011), 052,  \texttt{1107.2925}, 18 pages/ v2:
  argument improved, references added.

\bibitem{Marsh:2011aa}
D.~Marsh, L.~McAllister, and T.~Wrase, \emph{{The Wasteland of Random
  Supergravities}},  (2011),  \texttt{1112.3034}.

\bibitem{Danielsson:2011au}
U.~H. Danielsson, S.~S. Haque, P.~Koerber, G.~Shiu, T.~Van~Riet, et~al.,
  \emph{{De Sitter hunting in a classical landscape}}, Fortsch.Phys.
  \textbf{59} (2011), 897--933,  \texttt{1103.4858}.

\bibitem{Caviezel:2009tu}
C.~Caviezel, T.~Wrase, and M.~Zagermann, \emph{{Moduli Stabilization and
  Cosmology of Type IIB on SU(2)-Structure Orientifolds}}, JHEP \textbf{1004}
  (2010), 011,  \texttt{0912.3287}.

\bibitem{Dibitetto:2010rg}
G.~Dibitetto, R.~Linares, and D.~Roest, \emph{{Flux Compactifications, Gauge
  Algebras and De Sitter}}, Phys.Lett. \textbf{B688} (2010), 96--100,
  \texttt{1001.3982}.

\bibitem{Borghese:2011en}
A.~Borghese, R.~Linares, and D.~Roest, \emph{{Minimal Stability in Maximal
  Supergravity}},  (2011),  \texttt{1112.3939}, 27 pages, 1 figure.

\bibitem{Caviezel:2008tf}
C.~Caviezel, P.~Koerber, S.~Kors, D.~Lust, T.~Wrase, et~al., \emph{{On the
  Cosmology of Type IIA Compactifications on SU(3)-structure Manifolds}}, JHEP
  \textbf{0904} (2009), 010,  \texttt{0812.3551}.

\bibitem{Flauger:2008ad}
R.~Flauger, S.~Paban, D.~Robbins, and T.~Wrase, \emph{{Searching for slow-roll
  moduli inflation in massive type IIA supergravity with metric fluxes}},
  Phys.Rev. \textbf{D79} (2009), 086011,  \texttt{0812.3886}.

\bibitem{Dibitetto:2012ia}
G.~Dibitetto, A.~Guarino, and D.~Roest, \emph{{Exceptional Flux
  Compactifications}},  (2012),  \texttt{1202.0770}.

\bibitem{Danielsson:2012by}
U.~Danielsson and G.~Dibitetto, \emph{{On the distribution of stable de Sitter
  vacua}},  (2012),  \texttt{1212.4984}.

\bibitem{Danielsson:2012et}
U.~H. Danielsson, G.~Shiu, T.~Van~Riet, and T.~Wrase, \emph{{A note on
  obstinate tachyons in classical dS solutions}},  (2012),  \texttt{1212.5178}.

\bibitem{Blaaback:2013ht}
J.~Blaaback, U.~Danielsson, and G.~Dibitetto, \emph{{Fully stable dS vacua from
  generalised fluxes}},  (2013),  \texttt{1301.7073}.

\bibitem{Dibitetto:2011gm}
G.~Dibitetto, A.~Guarino, and D.~Roest, \emph{{Charting the landscape of N=4
  flux compactifications}}, JHEP \textbf{1103} (2011), 137,
  \texttt{1102.0239}.

\bibitem{Damian:2013dq}
C.~Damian, L.~R. Diaz-Barron, O.~Loaiza-Brito, and M.~Sabido, \emph{{Slow-Roll
  Inflation in Non-geometric Flux Compactification}},  (2013),
  \texttt{1302.0529}.

\bibitem{Wecht:2007wu}
B.~Wecht, \emph{{Lectures on Nongeometric Flux Compactifications}},
  Class.Quant.Grav. \textbf{24} (2007), S773--S794,  \texttt{0708.3984}.

\bibitem{Aldazabal:2006up}
G.~Aldazabal, P.~G. Camara, A.~Font, and L.~Ibanez, \emph{{More dual fluxes and
  moduli fixing}}, JHEP \textbf{0605} (2006), 070,  \texttt{hep-th/0602089}.

\bibitem{Frey:2002hf}
A.~R. Frey and J.~Polchinski, \emph{{N=3 warped compactifications}}, Phys.Rev.
  \textbf{D65} (2002), 126009,  \texttt{hep-th/0201029}.

\bibitem{Kachru:2002he}
S.~Kachru, M.~B. Schulz, and S.~Trivedi, \emph{{Moduli stabilization from
  fluxes in a simple IIB orientifold}}, JHEP \textbf{0310} (2003), 007,
  \texttt{hep-th/0201028}.

\bibitem{Flournoy:2004vn}
A.~Flournoy, B.~Wecht, and B.~Williams, \emph{{Constructing nongeometric vacua
  in string theory}}, Nucl.Phys. \textbf{B706} (2005), 127--149,
  \texttt{hep-th/0404217}.

\bibitem{Shelton:2005cf}
J.~Shelton, W.~Taylor, and B.~Wecht, \emph{{Nongeometric flux
  compactifications}}, JHEP \textbf{0510} (2005), 085,
  \texttt{hep-th/0508133}.

\bibitem{Shelton:2006fd}
J.~Shelton, W.~Taylor, and B.~Wecht, \emph{{Generalized Flux Vacua}}, JHEP
  \textbf{0702} (2007), 095,  \texttt{hep-th/0607015}.

\bibitem{Andriot:2011uh}
D.~Andriot, M.~Larfors, D.~Lust, and P.~Patalong, \emph{{A ten-dimensional
  action for non-geometric fluxes}}, JHEP \textbf{1109} (2011), 134,
  \texttt{1106.4015}.

\bibitem{Andriot:2010ju}
D.~Andriot, E.~Goi, R.~Minasian, and M.~Petrini, \emph{{Supersymmetry breaking
  branes on solvmanifolds and de Sitter vacua in string theory}}, JHEP
  \textbf{1105} (2011), 028,  \texttt{1003.3774}.

\bibitem{Geissbuhler:2013uka}
D.~Geissbuhler, D.~Marques, C.~Nunez, and V.~Penas, \emph{{Exploring Double
  Field Theory}},  (2013),  \texttt{1304.1472}.

\bibitem{deCarlos:2009fq}
B.~de~Carlos, A.~Guarino, and J.~M. Moreno, \emph{{Flux moduli stabilisation,
  Supergravity algebras and no-go theorems}}, JHEP \textbf{1001} (2010), 012,
  \texttt{0907.5580}.

\bibitem{deCarlos:2009qm}
B.~de~Carlos, A.~Guarino, and J.~M. Moreno, \emph{{Complete classification of
  Minkowski vacua in generalised flux models}}, JHEP \textbf{1002} (2010), 076,
   \texttt{0911.2876}.

\bibitem{Borghese:2012yu}
A.~Borghese, D.~Roest, and I.~Zavala, \emph{{A Geometric bound on F-term
  inflation}}, JHEP \textbf{1209} (2012), 021,  \texttt{1203.2909}.

\bibitem{Abbott:1981ff}
L.~Abbott and S.~Deser, \emph{{Stability of Gravity with a Cosmological
  Constant}}, Nucl.Phys. \textbf{B195} (1982), 76.

\end{thebibliography}
\addcontentsline{toc}{section}{Bibliography}
\bibliographystyle{TitleAndArxiv}

\end{document}